\documentclass[pre,twocolumn,aps,showpacs,superscriptaddress]{revtex4-2}
\usepackage{bm}%
\usepackage{float}
\usepackage[colorlinks=true,linkcolor=blue]{hyperref}%
\expandafter\ifx\csname package@font\endcsname\relax\else
 \expandafter\expandafter
 \expandafter\usepackage
 \expandafter\expandafter
 \expandafter{\csname package@font\endcsname}%
\fi

\usepackage{array}
\usepackage{amsmath,amssymb}
\usepackage{MnSymbol}
\usepackage{tikz-feynman} 
\usepackage[normalem]{ulem}
\usepackage{graphicx}
\usepackage{mathrsfs}
\usepackage{bm}
\usepackage{mathtools}
\usepackage{epstopdf}
\usepackage{hyperref}

\hypersetup{%
   pdfpagemode=UseNone, 
   pdfstartpage=1,
   pdfmenubar=true,
   pdftoolbar=true,
   colorlinks = true,
   linkcolor=blue,
   citecolor=blue,
   urlcolor=blue,
   bookmarksopen=false
 }
 \usepackage{amsmath}
\usepackage{amsfonts}
\usepackage{mathtools}
\usepackage{graphicx}
\usepackage{fixme}
\usepackage{color}
\usepackage{soul}
\usepackage{mhchem,chemformula}

\begin{document}
\title{Quantum multifractality as a probe of  phase space in the Dicke model}
\author{M. A.~Bastarrachea-Magnani}
\affiliation{Departamento de F\'isica, Universidad Aut\'onoma Metropolitana-Iztapalapa, Av. Ferrocarril San Rafael Atlixco 186, C.P. 09310 Mexico City, Mexico}
\author{D. Villase\~nor}
\affiliation{Instituto de Investigaciones en Matem\'aticas Aplicadas y en Sistemas, Universidad Nacional Aut\'onoma de M\'exico, C.P. 04510 Mexico City, Mexico}
\author{J. Ch\'avez-Carlos} 
\affiliation{Department of Physics, University of Connecticut, Storrs, Connecticut 06269, USA}
\author{S.  Lerma-Hern\'andez}
\affiliation{Facultad  de F\'\i sica, Universidad Veracruzana,  Campus Arco Sur, Paseo 112, C.P. 91097  Xalapa, Mexico}
\author{L. F. Santos} 
\affiliation{Department of Physics, University of Connecticut, Storrs, Connecticut 06269, USA}
\author{J. G. Hirsch} 
\affiliation{Instituto de Ciencias Nucleares, Universidad Nacional Aut\'onoma de M\'exico, Apdo. Postal 70-543, C.P. 04510 Mexico City, Mexico}

\begin{abstract}
We study the multifractal behavior of coherent states projected in the energy eigenbasis of the spin-boson Dicke Hamiltonian, a paradigmatic model describing the collective interaction between a single bosonic mode and a set of two-level systems. By examining the linear approximation and parabolic correction to the mass exponents, we find ergodic and multifractal coherent states and show that they reflect details of the structure of the classical phase space, including chaos, regularity, and features of localization. The analysis of multifractality stands as a sensitive tool to detect changes and structures in phase space, complementary to classical tools to investigate it. We also address the difficulties involved in the multifractal analyses of systems with unbounded Hilbert spaces.
\end{abstract}

\maketitle

\section{Introduction}
\label{sec:1} 

Mandelbrot first introduced the notion of multifractality to describe the statistical properties of turbulent flows~\cite{Mandelbrot1974}. Multifractality is characterized by an infinite set of critical exponents that determine the scaling of the moments of the distribution of some quantity. It has been observed in a wide range of phenomena, from mathematical objects, such as strange attractors~\cite{Grassberger1983a,Grassberger1984,Grebogi1988} and diffusion-limited aggregates~\cite{Hanan2012}, to the human heartbeat series~\cite{Ivanov1999} and brain activity~\cite{LaRocca2018}. In the last two decades, the subject has gained attention in the quantum domain, for both disordered and clean systems, in relation to problems, such as the Anderson localization phenomena~\cite{Mirlin2006,Vazquez2008,Rodriguez2008,Rodriguez2011}, many-body localization~\cite{DeLuca2013,Torres2015,Monthus2016,Solorzano2021}, quantum phase transitions~\cite{Lindinger2019}, disordered Josephson junctions~\cite{Pino2017}, the Bose-Josephson junction~\cite{Sinha2020}, Floquet eigenstates~\cite{Roy2018,Sarkar2021,Sarkar2022}, quantum phases in spin chains~\cite{Luitz2014}, quantum maps~\cite{Martin2010}, robustness against perturbations~\cite{Dubertrand2014,Dubertrand2015}, open quantum systems~\cite{Bilen2019}, quantum scarring~\cite{Meenakshisundaram2005}, and applications to quantum computing~\cite{Smelyanskiy2020}.

In quantum mechanics, multifractality roughly means that the wave function is extended but effectively restricted to a portion of the Hilbert space~\cite{Mirlin2000,Nakayama2003}.  This restriction happens because the weight of each component of the wave function scales differently and independently when the Hilbert space dimension increases. Each weight is a fractal on its own, hence the name multifractal. Quantum multifractality is intertwined with the concepts of localization, ergodicity, and chaos~\cite{Haake2018} and was recently used as a local measure of chaos for the kicked top model~\cite{Wang2021}. In this work, we perform quantum multifractal analyses to examine and compare the classically chaotic and regular structures of the phase space of the Dicke model, in the same spirit as what was done in Ref.~\cite{Bastarrachea2016}, where a measure of quantum state localization, the so-called participation ratio, was used to probe classical chaos in that model.

The spin-boson Dicke model describes a bosonic field strongly interacting with the collective degrees of freedom of $\mathcal{N}$ two-level systems (qubits)~\cite{Dicke1954}. It has drawn attention in recent years not only because it is the most simple, yet nontrivial, interacting model for exploring equilibrium and nonequilibrium properties~\cite{Garraway2011, Kirton2019,Larson2021,Chelpanova2023}, but also because it can be realized in various experimental setups, such as neutral atoms~\cite{Baumann2010,Baumann2011,Ritsch2013,Klinder2015}, ion traps~\cite{Cohn2018,Safavi2018}, and Raman cavities~\cite{Baden2014,Zhang2018}. One of the most prominent features of the model is the prediction of the transition to a superradiant quantum phase. In addition, the system's spectrum exhibits a transition from regularity to chaos as the energy increases~\cite{Perez-Fernandez2011,Chavez2016,Buijsman2017}, thus granting a fertile ground for exploring the onset of (multi-)fractality. 

Multifractality was recently studied in an interacting Tavis-Cummings model~\cite{Mattiotti2023} (an integrable version of the Dicke model without the counter-rotating terms), where the eigenstates written in the computational basis were shown to be nonergodic.  Fractality was also identified in the ground state of both the standard and anisotropic Dicke models~\cite{Das2022,Das2023}. However, investigating multifractality for energies above the ground state is challenging, because the model has an unbounded Hilbert space, and the study of multifractality relies on scaling analysis.

Here, we explore the multifractal behavior of coherent states spanned by the eigenbasis of the Dicke Hamiltonian. Each coherent state represents a point in the phase space where it is centered.  We show that the analysis of quantum multifractality can be used as a probe to identify chaos and regularity.

The paper is organized as follows. In Sec.~\ref{sec:2}, we describe the formalism for classical and quantum multifractality. In Sec.~\ref{sec:3}, we present the Dicke Hamiltonian, its classical limit, and the general protocol to study the multifractality of coherent states. In Sec.~\ref{sec:4}, we analyze the fractal dimension of representative coherent states, and in Sec.~\ref{sec:5}, we explore the overall behavior of multifractality over phase space structures across different energy surfaces. Finally, in Sec.~\ref{sec:6}, we offer our perspectives and conclusions. We also include appendices with further technical details. 

\section{Fractal Analysis}
\label{sec:2}

Generally speaking, multifractality is a tool for characterizing, in a statistical sense, the nature of a local positive measure, that is, how a positive quantity is distributed on a set supporting that measure~\cite{Paladin1987,Godreche1990}. The scaling of the measure provides information about the measure's singular local behavior. 

\subsection{Multifractality quantification}

In mathematics, to describe the dimensionality of an object quantitatively, one divides it into $N$ pieces labeled by $k=1,\ldots, N$, and considers an event occurring upon the object at the piece $k$ with a probability $p_{k}$, which is given by the measure $d\mu$ or resolution length~\cite{Paladin1987}. Next, one defines the partition function,
\begin{gather}
Z(q)=\sum_{k=1}^{N}p_{k}^{q}, 
\end{gather}  
where $Z(1)=1$ to fulfill normalization. The partition function provides global, quantitative information on the local behavior of the measure around each piece. As $N\rightarrow \infty$, the size of each piece decreases as $N^{-1}$, and the scaling behavior of the partition function is given by $Z(q) \sim N^{-\tau_{q}}$, where $\tau_{q}$ denotes the scaling in terms of $q$ only, and $\tau_{q=1}=0$ due to normalization. The exponents $\tau_{q}$ are called mass or homogeneity exponents~\cite{Ott2002}.

The measure is multifractal when $\tau_{q}$ is a nonlinear function of $q$. This is expressed by parametrizing the mass exponents as 
\begin{gather}
\label{eq:mass_exponent}
\tau_{q}=D_{q}(q-1),
\end{gather}
where $D_{q}$ is the generalized (R\'enyi) dimension~\cite{Hentschel1983,Halsey1986} defined as 
\begin{gather}
D_{q}=\lim_{N\to\infty} \left(\frac{1}{1-q}\frac{\log Z(q)}{\log N}\right).
\end{gather}
For $q=0$, the partition function counts the number of nonempty pieces of size $N^{-1}$, so  $\tau_{0}=-D_{0}=-D$, where $D$ is called Hausdorff fractal dimension or capacity of the support of the measure~\cite{Ott2002}. When $D_{q}$ is a constant function of $q$, the system is monofractal with dimension $D$. This also includes objects of integer dimension (typically considered nonfractals). The generalized dimensions have specific names for some values of $q$. $D_1$ is called information dimension, because it measures the information gained by observing a system's trajectory with some precision and quantifying the Kolmogorov-Sinai entropy after a long time of observation~\cite{Balatoni1956}. $D_2$ is known as the correlation dimension of the measure~\cite{Grassberger1983a,Grassberger1984}, because it corresponds to the scaling of the correlation between two points in the classical phase space. It has been conjectured that in the case of strange attractors, $D_{1}$ and $D_{2}$ are related to the Lyapunov exponents~\cite{Grassberger1983b}.

The mass exponents and the generalized dimensions follow some rules that need to be observed~\cite{Ott2002}. The exponents $\tau_{q}$ must be monotonically increasing functions with negative curvature~\cite{Halsey1986}. Hence, it holds that $d\tau_{q}/dq>0$ and $d^{2}\tau_{q}/dq^{2}\leq 0$. Instead, the generalized dimensions $D_{q}$ are positive monotonically decreasing functions of $q$ bounded by $D_{\pm\infty}=D(q\rightarrow\pm\infty)$~\cite{Janssen1994}. Thus, $d D_{q}/dq\leq 0$, and $0\leq D_{\infty}\leq D_{q}\leq D_{-\infty}$. 

\subsection{Quantum multifractality}

One way to bring the concept of multifractality to the quantum realm is in terms of the level of delocalization of a quantum state written in a given basis. Consider a quantum state $|\Psi\rangle$ and a complete basis $\{|k\rangle\}$ over a Hilbert space of dimension $\aleph$. The probability of finding the quantum state in one of the elements of the basis is given by $|c_{k}|^{2}=|\langle k|\Psi\rangle|^{2}$, which plays the role of the probability of finding the state (the event) over a given eigenstate (the piece). Thus, in this case, the partition function can be built in terms of the generalized inverse participation ratios~\cite{Evers2008} as 
\begin{gather}
Z(q)=\text{IPR}_{q}=\sum_{k=1}^{\aleph}|c_{k}|^{2q},
\end{gather}
Here, the defining integrated measure is the normalization condition, so $\text{IPR}_{q=1}=1$. The case $q=2$ corresponds to the standard inverse participation ratio (IPR)~\cite{Bell1970,Thouless1974}. The scaling of $\text{IPR}_{q}$ with the size of the Hilbert space reveals the asymptotic statistics of the participation of the basis elements $|k\rangle$ in the state $|\Psi\rangle$ and is expected to be of the form $\text{IPR}_{q}\sim \aleph^{-\tau_{q}}$, where the exponents $\tau_{q}$ are now defined as
\begin{equation} \label{eq:tau}
\tau_{q}:=-\lim_{\aleph\rightarrow\infty}\frac{\log \text{IPR}_{q}}{\log(\aleph)}.
\end{equation}
The scaling of $\text{IPR}_{q}$ for all $q$ is used to classify the states. Localized states have $D_{q}=0$, extended but nonergodic states have $0\leq D_{q}<1$, and ergodic states imply $D_{q}=1$. Multifractal wave functions are nonergodic extended states, because the ratio between the effective portion of the Hilbert space that they occupy and the full size of the Hilbert space is neither one nor vanishing as the system size increases~\cite{Lindinger2019}. In condensed matter, systems classified as insulators have $D_{q}=0$, and conductors have $D_{q}\neq 0$~\cite{Evers2008}. However, a full theory linking specific nonlinear behaviors of either $\tau_{q}$ or $D_{q}$ to specific physical phenomena for all values of $q$ is, in general, absent. Depending on the system, the connection is usually made ad hoc. 

To determine whether $\tau_{q}$ is nonlinear, one  resorts to the so-called anomalous scaling exponent,
\begin{gather}
\Delta_{q} = \tau_{q} - D(q-1),
\label{Eq:Deltaq}
\end{gather}
which quantifies the deviation of the mass exponents from the linear behavior~\cite{Evers2008}. In the above equation, $D$ is the linear approximation to $D_{q}$. If the behavior of the mass exponents with $q$ is parabolic and the anomalous fractal dimension is quadratic, $\Delta_{q}\approx \Delta (1-q)q$, with $\Delta$ constant, then we have {\it weak multifractality}, a concept introduced as an effort to provide a first approximation for multifractal behavior. Some symmetries have been proved for the anomalous fractal dimension, such as the reciprocity relation $\Delta_{q}=\Delta_{1-q}$~\cite{Mirlin2006,Monthus2009,Bogomolny2012}, that has been predicted to be valid for transitions belonging to the Wigner-Dyson classes. Efforts have been made to show its universality, being proved in some cases including the critical point of the Anderson localization transition~\cite{Mirlin2006,Mildenberger2007b,Rodriguez2009,Obuse2010}, the power-law banded random matrix~\cite{Mildenberger2007}, and the integral quantum Hall effect~\cite{Evers2008b}. Conversely, it has been found that the relation is not fulfilled when mechanisms like Gaussian fluctuations at small scales or algebraic localization of the wave function are present, such as in Floquet critical systems and random graphs~\cite{Bilen2021}. However, as is  discussed below, given numerical convergence constraints, we cannot verify this relation systematically for the Dicke model.

In what follows, we investigate linear and quadratic fits for $\tau_{q}$ as a function of $q$ and then use Eq.~\eqref{Eq:Deltaq} to obtain the approximate values to $D$ and $\Delta$. If the linear fit of $\tau_{q}$ is very good, then $\Delta\simeq 0$, so the system is closer to a monofractal.  

\section{Dicke Hamiltonian}
\label{sec:3}

The spin-boson Dicke Hamiltonian is given by 
\begin{equation}
\label{eqn:qua_hamiltonian}
\hat{H}_{\text{D}}=\omega\hat{a}^{\dagger}\hat{a}+\omega_{0}\hat{J}_{z}+\frac{\Omega}{\sqrt{\mathcal{N}}}(\hat{a}^{\dagger}+\hat{a})\hat{J}_{x},
\end{equation}
where $\hbar=1$, $\hat{a}^{\dagger}$ ($\hat{a}$) is the bosonic creation (annihilation) operator, and $\hat{J}_{x,y,z}=\frac{1}{2}\sum_{i=1}^{\mathcal{N}}\hat{\sigma}_{x,y,z}^{i}$ are collective pseudospin operators satisfying the su(2) algebra, with each Pauli matrix $\hat{\sigma}^i_{x,y,z}$  describing a single spin-1/2 (qubit). Here, $\omega$, $\omega_0$, and $\Omega$ are the boson, single qubit, and Rabi energy splittings. The Hamiltonian has two symmetries. First, it commutes with the total pseudospin length operator $\hat{\textbf{J}}^{2}=\hat{J}_{x}^{2}+\hat{J}_{y}^{2}+\hat{J}_{z}^{2}$, so the Hilbert space is divided into different subspaces corresponding to the pseudospin length $j$. The ground state of the collective system lies on the totally symmetric subspace with maximum pseudospin length $j=\mathcal{N}/2$, where the collective degrees of freedom are equivalent to bosons. Second, the Hamiltonian commutes with a discrete parity operator, $\hat{\Pi}=\exp\left[i \pi (\hat{a}^{\dagger} \hat{a}+ \hat{J}_z+j\hat{\mathbb{I}})\right]$, which leads to the further separation of the Hilbert space into two subspaces corresponding to each parity value.

An advantage of working within the totally symmetric subspace, as done here, is that one can straightforwardly associate a classical Hamiltonian to the Dicke model in Eq.~\eqref{eqn:qua_hamiltonian} by employing coherent states. The corresponding classical Hamiltonian $h_{\text{D}}$ is obtained by taking the expectation value of $\hat{H}_{\text{D}}$ under the tensor product of bosonic Glauber $|\beta\rangle$ and atomic Bloch $|w\rangle$ coherent states~\cite{Deaguiar1991,Deaguiar1992,Bastarrachea2014a,Bastarrachea2014b,Bastarrachea2015,Chavez2016,Villasenor2020}
\begin{gather}
|\bm{z}\rangle=|\beta\rangle\otimes|w\rangle=\frac{e^{-|\beta|^2/2}}{(1+|w|^2)^{j}}e^{\beta\hat{a}^{\dagger}}e^{w\hat{J}_{+}}|0\rangle\otimes|j,-j\rangle,
\end{gather}
where $|0\rangle$ and  $|j,-j\rangle$ are the boson and pseudospin fiducial states, respectively~\cite{Gilmore1990}. By dividing over $j$, we obtain
\begin{gather}
h_\text{D}(\bm z) = j^{-1}\langle\bm{z}|\hat{H}_{\text{D}}|\bm{z}\rangle=j^{-1}\omega|\beta|^{2}-\omega_{0}\left(\frac{1-|w|^{2}}{1+|w|^{2}}\right)\\ \nonumber
+\frac{\Omega\left(\beta+\beta^{*}\right)\left(w+w^{*}\right)}{\sqrt{2j}(1+|w|^{2})}.
\end{gather}
Changing to canonical variables $(x,p)$ and $(\phi,j_{z})$ in phase space, it reads
\begin{gather}
h_\text{D}(\bm z) =\frac{\omega}{2}\left(x^{2}+p^{2}\right)+\omega_{0}j_{z}+\Omega\sqrt{1-j_z^{2}} x\cos\phi,
\end{gather}
where $\beta=\sqrt{j/2}\left(x+ip\right)$, $w=\tan(\theta/2)e^{-i\phi}$, $j_z=-\cos\theta$, and $\phi=\tan^{-1}(j_y/j_x)$. While $x$ and $p$ are the classical quadratures of the field, $\theta$ and $\phi$ are the spherical angles of the angular momentum vector $\vec{j}=j(j_x,j_y,j_z)$ lying in the Bloch sphere. Using coherent states, one can establish a direct quantum-classical correspondence, where a given coherent state $|\bm{z}_{0}\rangle=|x_{0},p_{0};\phi_{0},j_{z0}\rangle$ represents a point in the classical phase space. We exploit this connection to explore the multifractal properties of the classical phase space from a quantum point of view, much in the sense of what was done for $D_2$ in Ref.~\cite{Bastarrachea2016}. 

We work in the parameter space of the superradiant phase. The superradiant quantum phase transition occurs when the Rabi splitting attains the critical value $\Omega_{c}=\sqrt{\omega\omega_{0}}$, and the system goes from an uncorrelated, normal phase ($\Omega<\Omega_c$) to a strongly correlated, superradiant phase ($\Omega>\Omega_c$) ~\cite{Hepp1973a,Hepp1973b,Wang1973,Emary2003}. It has been shown that in the superradiant phase, the spectrum of the model exhibits a transition from regularity to chaos as the energy increases for both the quantum and classical Hamiltonians~\cite{Bastarrachea2015,Chavez2016}. We set the Rabi splitting to $\Omega=2.0\Omega_{c}$, where the regular-to-chaos transition happens smoothly.

\begin{figure}
\centering
\begin{tabular}{c}
\includegraphics[width=0.5\textwidth]{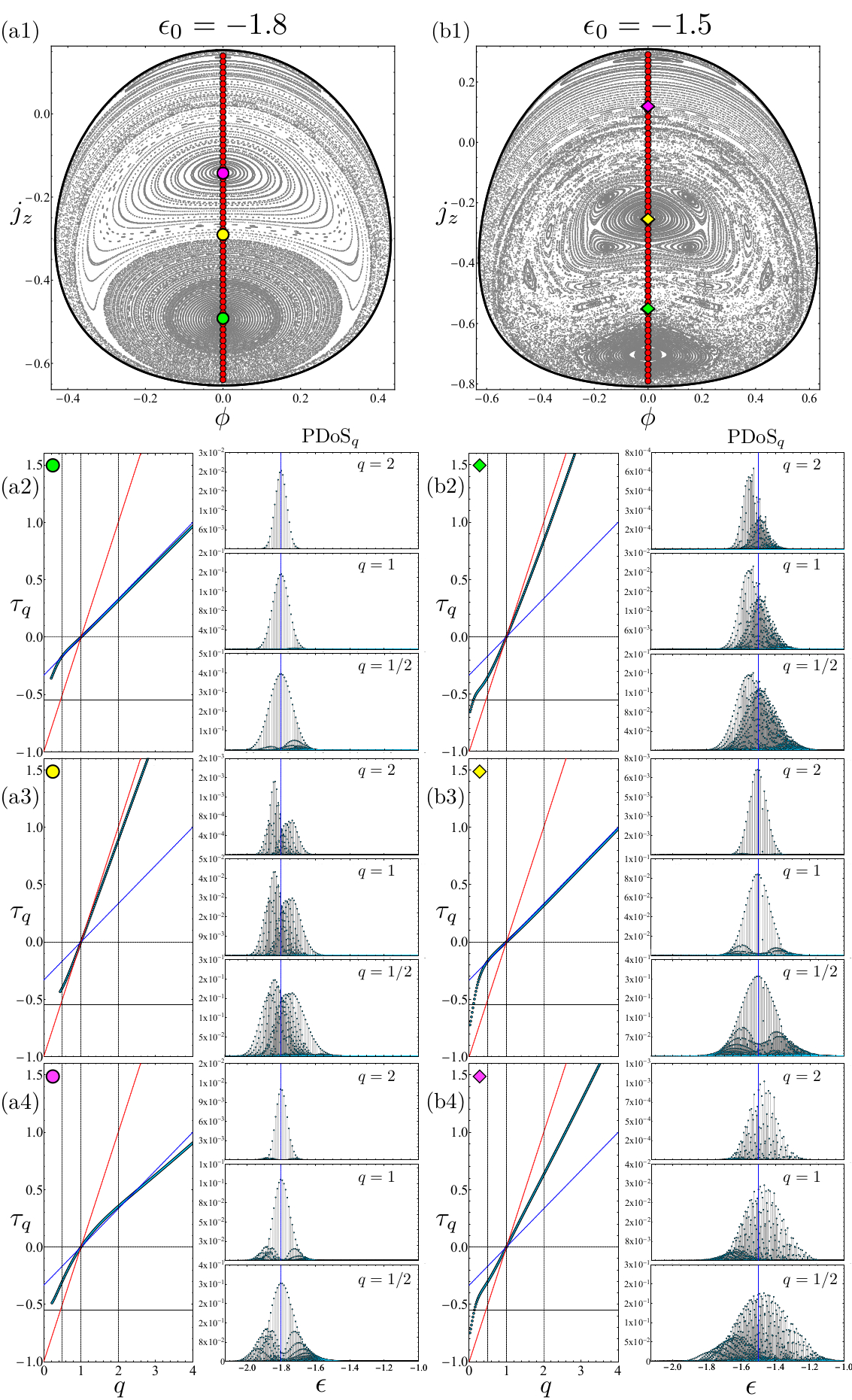}
\end{tabular} 
\caption{(Top row) Poincar\'e surface of sections for  two selected energies, $\epsilon_{0}=-1.8$ (a1) and $\epsilon_{0}=-1.5$ (b1), where there is predominance of regular structures in phase space. (Below) Mass exponents $\tau_q$ (first and third columns) as a function of $q$ for selected coherent states in each energy surface, and their PDoS$_{q}$ (second and fourth columns) for different values of $q$, and $\epsilon_{0}=-1.8$ (a2-a4) or $\epsilon_{0}=-1.5$ (b2-b4); $j=100$ for all panels. The location of each coherent state in the Poincar\'e surface is marked with a symbol (circles for $\epsilon_{0}=-1.8$ and diamonds for $\epsilon_{0}=-1.5$). In the $\tau_{q}$ plots, the dotted cyan line corresponds to numerical results, the colored lines indicate the ergodic (red) and regular (blue) limiting cases of the $D_{q}$, the vertical black solid lines  mark $q=1/2,1$, and $2$. In the PDoS$_{q}$ plots, the vertical blue line marks the mean energy.}
\label{fig:1}
\end{figure}

\begin{figure}
\centering
\begin{tabular}{c}
\includegraphics[width=0.5\textwidth]{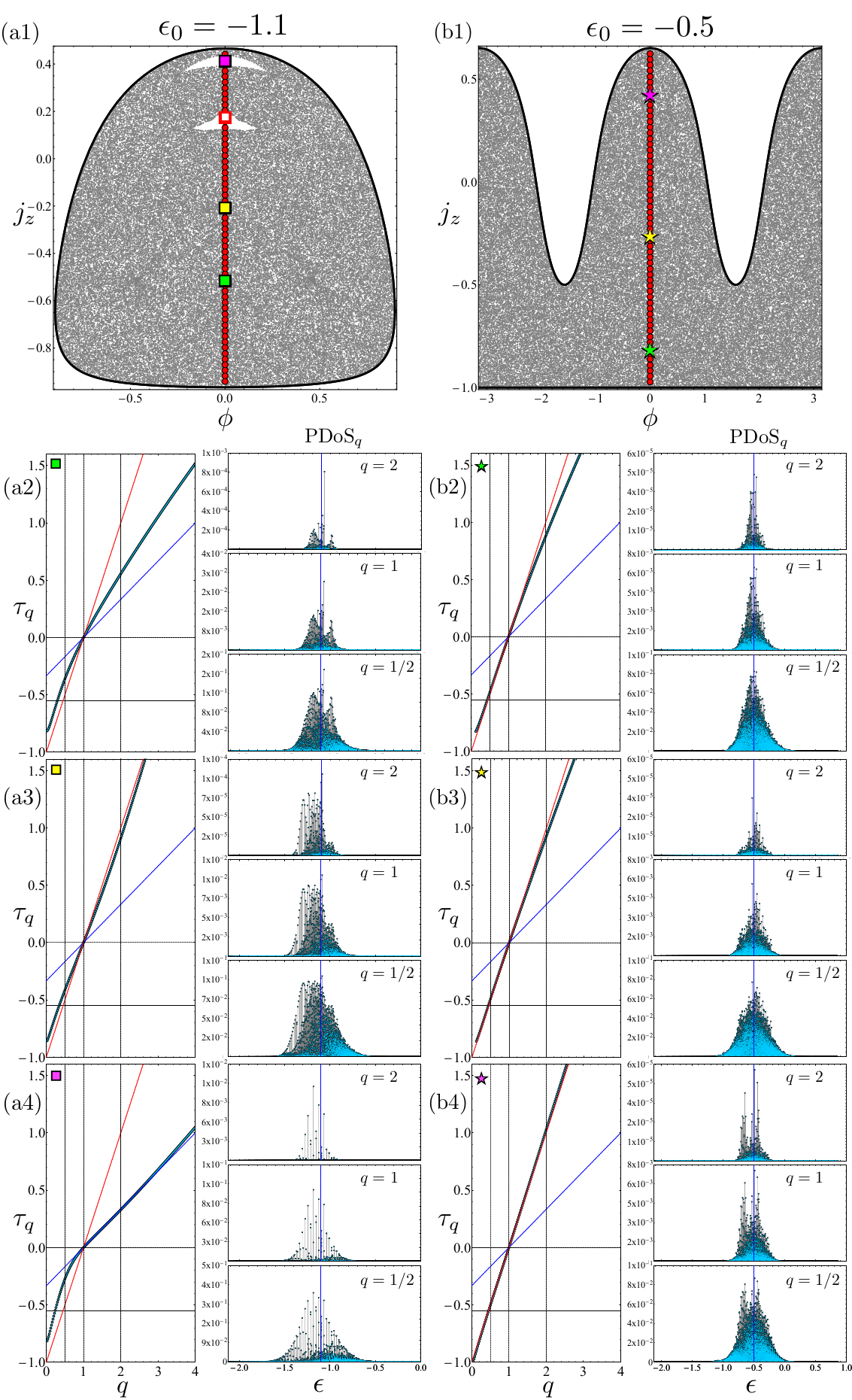}
\end{tabular} 
\caption{Similar to Fig.~\ref{fig:1} for the two selected energies $\epsilon_{0}=-1.1$ (a1) and $\epsilon_{0}=-0.5$ (b1), where there is predominance of chaos. Notice the change of scale in panel (b1) due to the sudden increase of available phase space at high energies~\cite{Bastarrachea2014a}. The location of each coherent state in the Poincar\'e surface is marked with a symbol (squares for $\epsilon=-1.1$ and stars for $\epsilon=-0.5$).}
\label{fig:2}
\end{figure}

\section{Multifractality in the Dicke model}
\label{sec:4}  

The Dicke model is nonintegrable, because it does not possess enough conserved quantities as degrees of freedom, so it must be solved numerically. Due to the bosonic degree of freedom, the Hilbert space is unbounded, which constitutes a significant challenge for the complete analysis of multifractality. To obtain the energy spectrum, one has to impose a cut-off to the bosonic subspace (irrespective of the chosen basis) and then ensure that the eigenfunctions are converged up to the desired energy that one plans to investigate~\cite{Bastarrachea2014PSa,Bastarrachea2014PSb}.

\subsection{Convergence and effective dimension}

We use a convergence criterion for $q=1$ to ensure that most of the wave function of the coherent states that we consider lie within the energy interval of numerically converged eigenstates, ranging from the ground-state to a selected excited energy, so minor components over higher energies remain negligible, and the normalization is guaranteed to a set of significant figures. Nevertheless, minor components associated with high-lying states are magnified and become not negligible for $q\sim 0$. The analysis of multifractality is thus challenging for $\text{IPR}_{q}$ with $q\sim 0$, which are highly sensitive to the convergence of the wave functions and the truncation. In particular, the Hausdorff dimension $D_0=D$ of the coherent states can hardly ever be identified for the Dicke model, although it can be estimated for some cases. The convergence problem is minimized by increasing the value of the cut-off, but this becomes highly expensive in terms of computational resources. 

The scaling analysis also requires setting a finite, effective dimension $\aleph_{\text{eff}}$ for the coherent states with a given mean energy. A naive choice would be the size of the collective qubit space, which goes as $j=\mathcal{N}/2$. Nevertheless, by studying the participation of states in a given energy surface and considering that the density of states (DoS) scales as $j$, it was recently shown that the proper scaling should go as $\aleph_{\text{eff}}\sim j^{3/2}$, instead of $\aleph_{\text{eff}}\sim j$~\cite{Pilatowsky-Cameo2022}. Therefore, we use $\aleph_{\text{eff}}\sim j^{3/2}$ as the effective size of the Hilbert space to perform the multifractal analysis. 

\begin{figure}
\centering
\begin{tabular}{c}
\includegraphics[width=0.4\textwidth]{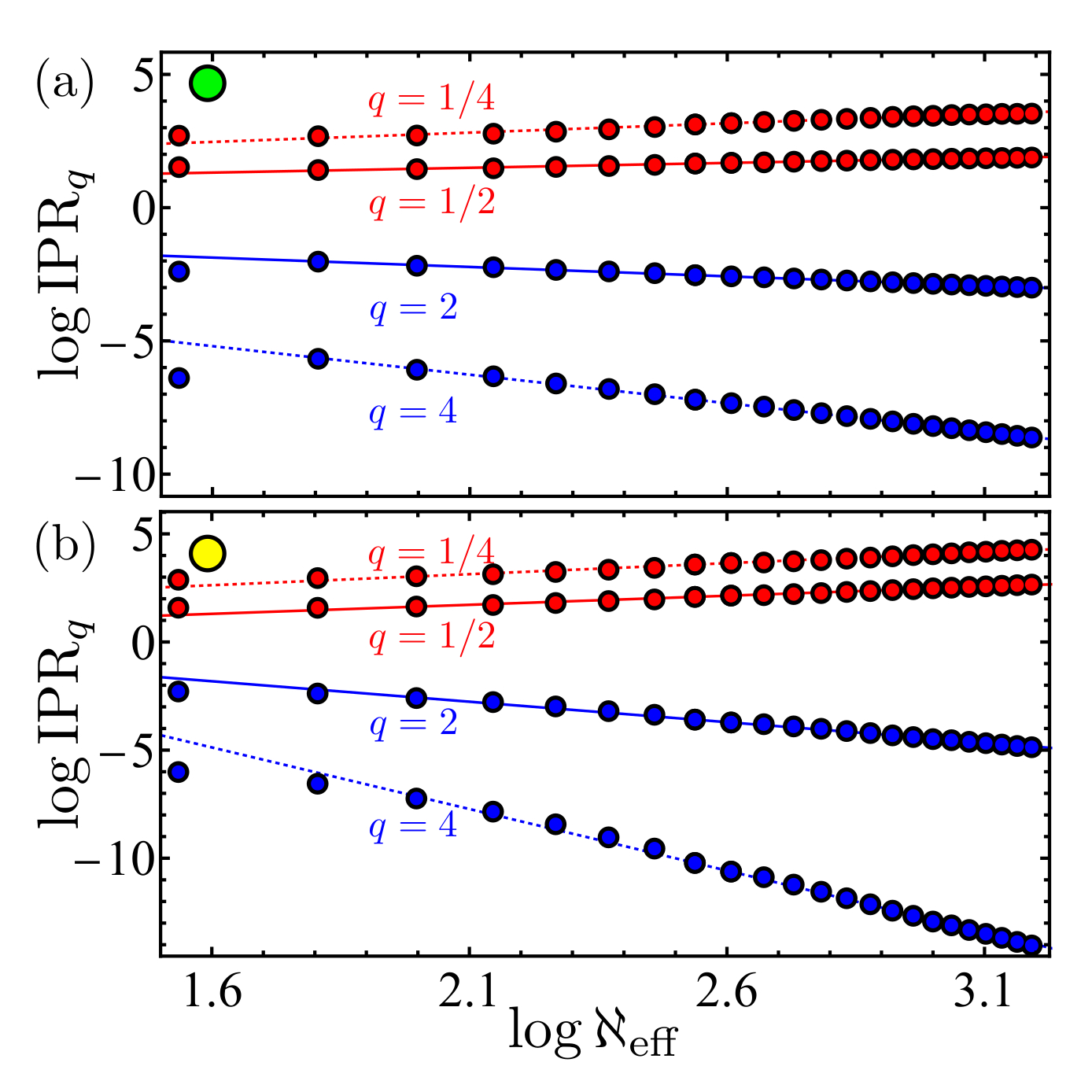}
\end{tabular} 
\caption{Logarithmic plot of $\text{IPR}_{q}$ as a function of $j^{3/2}$ ($j=5$ to $120$) for (a) the same coherent state as in Fig.~\ref{fig:1}~(a2) and (b) the same coherent state as in Fig.~\ref{fig:1}~(a3), for  $q=1/4,1/2,2$, and $4$. The linear fits ignore the first  points, from $j=5$ to $j=20$, to avoid finite-size effects. }
\label{fig:3}
\end{figure}

\begin{figure}
\centering
\begin{tabular}{c}
\includegraphics[width=0.4\textwidth]{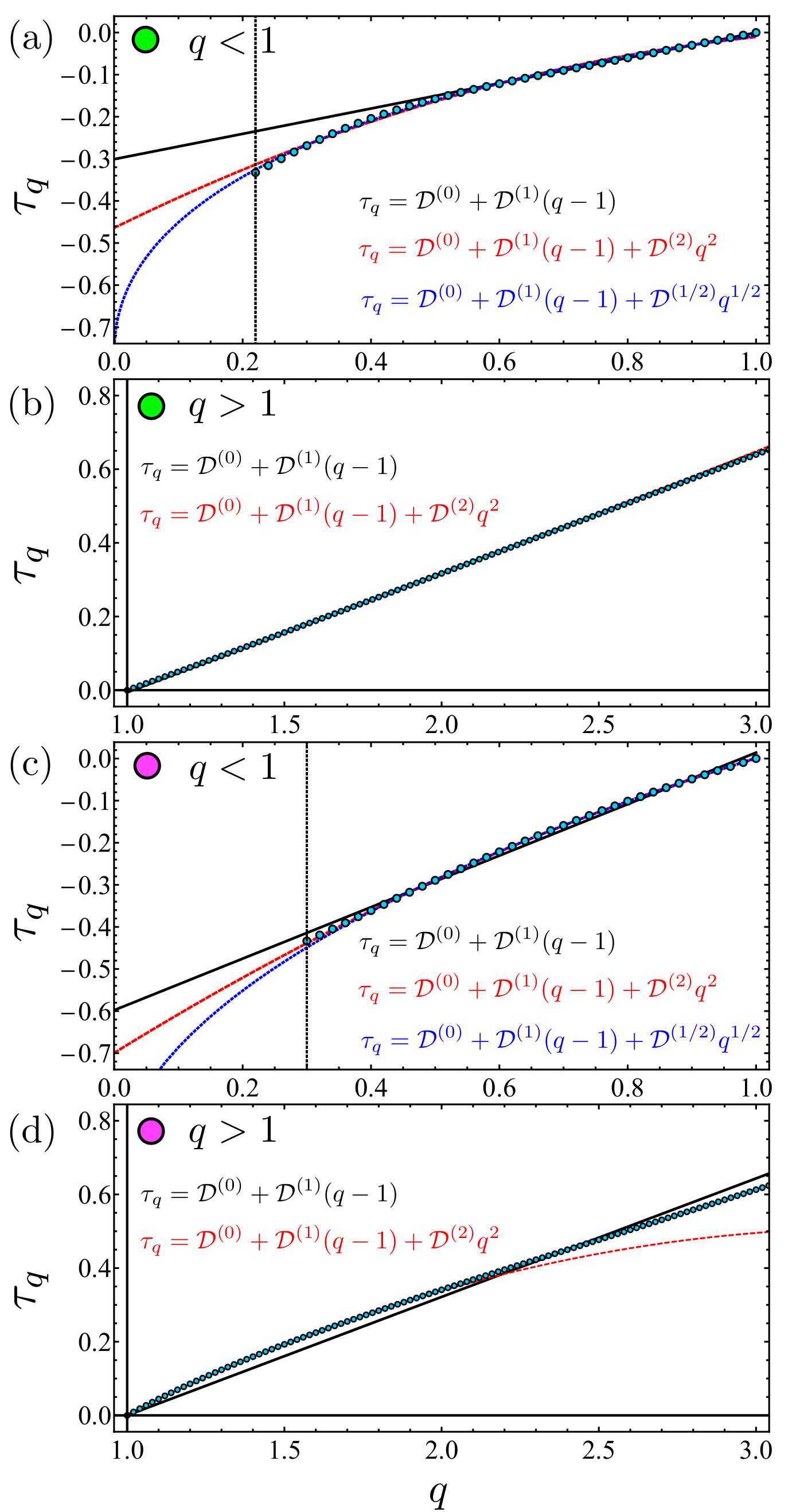}
\end{tabular} 
\caption{Mass exponent $\tau_{q}$ (blue dots) as a function of $q$ for the coherent state in Fig.~\ref{fig:1}~(a2) for (a) $q<1$ and (b) $q>1$, and for the coherent state in Fig.~\ref{fig:1}~(a4) for (c) $q<1$ and (d) $q>1$. Three different fits are shown: linear (solid black line), quadratic (red dashed line), squared root (blue dashed line). The dashed vertical black line represents the lowest value of $q$ where the convergence of the wave function is applicable.}
\label{fig:4}
\end{figure}

\subsection{Limiting cases of the generalized dimension}

To analyze the numerical results, we contrast them with two bounds for the scaling of $\text{IPR}_{q}$. They are obtained as follows.

(i) First, we consider a state $|\Psi^{(r)}\rangle$ with random components and a Gaussian profile. We can estimate the scaling of $\text{IPR}_{q}$ exactly (see Appendix~\ref{app:1}),
\begin{gather}
\text{IPR}_{q}^{(r)}\sim \aleph_{\text{eff}}^{(1-q)}.
\end{gather}
The mass exponent is linear as a function of $q$, $\tau_{q}^{(r)}=q-1$, and the state is described in terms of a single fractal of dimension $D_{q}^{(r)}=D=1$. This corresponds to an extended state and constitutes an overall upper bound of localization of the coherent states in the energy eigenbasis. This bound is indicated with a red dotted line in both Figs.~\ref{fig:1} and~\ref{fig:2}, panels~(a2)-(a4) and~(b2)-(b4). 

(ii) The other bound is obtained with a state $|\Psi^{(c)}\rangle$ for which most of the components in the eigenbasis are zero, except for a single sequence of nearly equally spaced energy levels $\{..., E_{k}^{(\text {seq})}, E_{k+1}^{(\text {seq})},...\}$. We assume that these nonzero components also follow a Gaussian profile. This would be the case of a coherent state centered in a point in phase space over a regular classical orbit with a single frequency $\omega_{\text{cl}}=E_{k+1}^{(\text {seq})}- E_{k}^{(\text {seq})}$~\cite{Lerma2018}. As it is shown in Appendix~\ref{app:1}, one obtains that 
\begin{gather}
\text{IPR}_{q}^{(c)}\sim \aleph_{\text{eff}}^{\frac{1}{3}(1-q)}.
\end{gather}
This means that $\tau_{q}^{(c)}=(1/3)(q-1)$ is also linear and that a single fractal of dimension $D_{q}^{(c)}=D=1/3$ can describe the state. This is an overall lower bound for the localization of a coherent state over the eigenbasis (lower bound for $D_q$). This case is indicated as a blue dotted line in both Figs.~\ref{fig:1} and~\ref{fig:2}, panels~(a2)-(a4) and~(b2)-(b4). We note that for a Gaussian distribution of a real state, where the components may be small but not exactly zero, the curve of $\tau_{q}$ versus $q$ in the region $0<q<1$ deviates from $D=1/3$. Then, for $q=0$, all the components have the same weight, and the distribution covers the full available support, so $D_{0}=-1$.

We expect the fractal dimension of most coherent states to be in the interval $D_{q}\in[D_{q}^{(c)}, D_{q}^{(r)}]$ when neglecting minor deviations due to finite-size effects and convergence of the wave functions. However, some states may have $D=0$ for $q>1$ thus presenting localization features for $q>1$.

\subsection{Procedure for the analysis of multifractality}

Our procedure to study the multifractal properties of coherent states goes as follows. We select a coherent state $|\bm{z}_{0}\rangle$ centered at the point $\bm{z}_{0}$ in phase space. We choose a set of points from an energy surface that coincides with one of the Poincar\'e surfaces of sections as shown in Figs.~\ref{fig:1} and~\ref{fig:2}, panels~(a1) and (b1). The Poincar\'e sections are chosen by setting $p_{0}=0$ and selecting the positive branch of the solution $x_{0}^{+}=x_{0}^{+}(\epsilon_0,p_0=0;\phi,j_z)$ that results from solving the equation $h_{\text{D}}(x_{0},p_{0}=0;\phi,j_{z})=\epsilon_0$ for a given energy $\epsilon_0=E_{0}/j$~\cite{Bastarrachea2016}. The points in phase space are thus located using only the atomic variables $(\phi,j_z)$. Next, we sample the energy surface by focusing only on the points along $\phi_{0}=0$ [red points in Figs.~\ref{fig:1} and~\ref{fig:2}, panels~(a1) and~(b1)]. 

The surface energies grant insight into different regions of chaos and regularity: for $\epsilon_0=-1.8$, there is a major presence of regular orbits with chaos emerging in small regions [Fig.~\ref{fig:1}~(a1)]; for $\epsilon_0=-1.5$ the phase space is mixed [Fig.~\ref{fig:1}~(b1)]; for $\epsilon_0=-1.1$, there is a predominance of chaos with some remaining regular islands [Fig.~\ref{fig:2}~(a1)]; and at $\epsilon_0=-0.5$, the system appears to be fully chaotic [Fig.~\ref{fig:2}~(b1)]. Within each energy surface, we choose a set of three representative points marked with colored symbols, whose coordinates $j_{z0}$ are given in Table~\ref{tab:1}. 

\begin{table}[h]
\begin{center}
\begin{tabular}{| c  c c c | c  c c c |} 
\hline
 \multicolumn{4}{|c|}{$\epsilon_0=-1.8$} &   \multicolumn{4}{c|}{$\epsilon_0=-1.5$}  \\ [0.5ex] 
 \hline
 & (a2) & (a3) & (a4) & & (b2) & (b3) & (b4) \\ 
  $j_{z0}$ & -0.492 & -0.290 & -0.143 &  $j_{z0}$ & -0.548 & -0.250 &  0.123  \\ [1ex]
 \hline
 \multicolumn{4}{|c|}{$\epsilon_0=-1.1$} &  \multicolumn{4}{c|}{$\epsilon_0=-0.5$} \\
 \hline
  & (a2) & (a3) & (a4) & & (b2) & (b3) & (b4)  \\ 
  $j_{z0}$ &-0.512 & -0.202 & 
  0.418 & $j_{z0}$ & -0.807 & -0.257 & 0.431 \\ [1ex]
 \hline
\end{tabular}
\caption{Values of $j_{z0}$ for twelve representative points in the Poincar\'e surfaces in Figs.~\ref{fig:1} and~\ref{fig:2}, panels~(a1) and (b1). The points are taken with $\phi_0=0$, $p_0=0$ and the positive branch of the solution for $x_{0}^{+}$ at the corresponding energy $\epsilon_{0}$.}
\label{tab:1}
\end{center}
\end{table}

Then, we solve the Dicke Hamiltonian numerically for several values of $j$ from $5$ to $120$ ($2j$ qubits) and a cut-off of $n_{\text{max}}=300$ quanta in the bosonic field to ensure convergence of about 30,000 eigenstates. See Appendix~\ref{app:3} for a comparison between the truncation for $n_{\text{max}}=300$ and $n_{\text{max}}=250$. The first step for our analysis of multifractality is to get the mass exponents. For each $j$ and each coherent state considered, we calculate $\text{IPR}_{q}$ and, according to Eq.~\eqref{eq:tau}, obtain $\tau_q$ by linearly fitting the logarithmic plot of $\text{IPR}_{q}$ versus $\aleph_{\text{eff}}$, as shown in Figs.~\ref{fig:3}~(a) and~\ref{fig:3}~(b) for two representative coherent states, as examples. To avoid finite-size effects, we ignore the first points (from $j=5$ to $j=20$) when performing the fittings in Fig.~\ref{fig:3}. The results obtained are used in the plots of $\tau_q$ versus $q$ shown in Figs.~\ref{fig:1}~(a2)-\ref{fig:1}~(a4), \ref{fig:1}~(b2)-\ref{fig:1}~(b4) and Figs.~\ref{fig:2}~(a2)-\ref{fig:2}~(a4). These plots are subsequently used to extract the generalized dimensions $D_q$. 
To assist our studies of multifractality, we also examine the distribution of the coefficients of each coherent state with respect to energy for a given $q$, what we call generalized participation local density of states, 
\begin{gather}
\text{PDoS}_{q}(\epsilon)=\sum_{k}|c_{k}|^{2q}\delta(\epsilon-\epsilon_{k}),
\end{gather}
where $|c_{k}|^{2q}=|\langle E_{k}|\bm{z}_{0}\rangle|^{2q}$, $|E_{k}\rangle$ are the eigenstates of the Dicke Hamiltonian, and $\epsilon_{k}=E_{k}/j$ are the scaled eigenergies. The PDoS$_q$ is related to the generalized local density of states~\cite{Zarate-Herrada2023} upon changing $2q$ by $q$. Plots for the PDoS$_{q}$ are shown in Figs.~\ref{fig:1} and~\ref{fig:2}. The analysis of PDoS$_{q}$ for different values of $q$ reveals structures hidden in the coherent states that help us better understand the source of multifractality. 

\subsection{Predominance of regularity }

We start by analyzing three representative coherent states within a low-energy domain ($\epsilon_{0}=-1.8$), where regularity is predominant in phase space. 

\subsubsection{Green circle in Fig.~\ref{fig:1}~(a1)}

The green circle in Fig.~\ref{fig:1}~(a1) indicates the center of a coherent state placed near a stable periodic orbit that emanates from a pseudo-conserved quantity of the Dicke Hamiltonian~\cite{Relano2016,Bastarrachea2017}. This coherent state is mainly described by a subset of nearly equally spaced eigenstates, and its PDoS$_{q}$ for $q=1$, shown in Fig.~\ref{fig:1}~(a2), has a Gaussian profile. The shape of PDoS$_{q}$ for $q>1$, as illustrated for $q=2$ also in Fig.~\ref{fig:1}~(a2), remains Gaussian, because small coefficients $|c_{k}|^{2q}$ become negligible as $q$ increases from 1. Accordingly, the $D_{q}$ for $q>1$, obtained using Eq.~\eqref{eq:mass_exponent} and the data in Fig.~\ref{fig:1}~(a2), follows almost exactly $D_{q}^{(c)}=1/3$.

The result for $D_q$ with $q>1$ suggests that the coherent state is monofractal. However, for $q<1$, new structures emerge within the Gaussian envelope of PDoS$_{q}$, as seen for $q=1/2$ in  Fig.~\ref{fig:1}~(a2). They are caused by small contributions $|c_{k}|^{2q}$ that are associated with classical trajectories slightly away from the center of the set of regular orbits in the Poincar\'e section, which are sampled by the selected coherent state. In turn, this is reflected in the behavior of the mass exponents as a function of $q$. For $q<1$, $\tau_{q}$ deviates from the linear behavior, thus suggesting multifractality. 

To determine whether the state is then mono- or multifractal, a better picture is achieved by analyzing $\tau_q$ versus $q$ for $q<1$ in Fig.~\ref{fig:4}~(a) and for $q>1$ in Fig.~\ref{fig:4}~(b). We show in Fig.~\ref{fig:4}~(a) that the linear fitting, given by $\tau_{q}=\mathcal{D}^{(0)}+\mathcal{D}^{(1)}(q-1)$ (solid black line), fails to describe the behavior of $\tau_q$ around $q\in[0.2,0.5]$, as it becomes nonlinear when $q$ decreases. Instead, both the parabolic fitting $\tau_{q}=\mathcal{D}^{(0)}+\mathcal{D}^{(1)}(q-1)+\mathcal{D}^{(2)}q^{2}$ (red dashed curve) and the square root $\tau_{q}=\mathcal{D}^{(0)}+\mathcal{D}^{(1)}(q-1)+\mathcal{D}^{(1/2)}q^{1/2}$ (blue dashed curve) perform better. Instead, for $q>1$ in Fig.~\ref{fig:4}~(b), the behavior is mostly linear, being described by the slope $D_{q}^{(c)}=1/3$. Thus, the result for $q<1$ makes it clear that this state is multifractal, and it serves to benchmark the multifractal behavior in the regular region. 
As we shall see, the nonlinear departure from $D_{q}^{(c)}$ for $q<1$ also appears for other coherent states centered in regular regions of the phase space. 

\subsubsection{Yellow circle in Fig.~\ref{fig:1}~(a1)}

The coherent state centered at the yellow circle in Fig.~\ref{fig:1}~(a1) is close to the stochastic layer where chaos is emerging~\cite{Zaslavskii1991}. In this case, the PDoS$_{q}$ for $q=1$ is made of a set of Gaussians. This feature is invariant and persists for larger or smaller values of $q$. As a result, the mass exponent in Fig.~\ref{fig:1}~(a3) is basically a straight line leading to a value of $D_q$ very close to the upper bound $D_{q}^{(r)}=1$. This implies that this state is not multifractal but an ergodic state.

At very small values of $q \ll 1$, there is a change in the curvature of the mass exponent [the beginning of this change is noticeable in Fig.~\ref{fig:1}~(a3)], but this is an artifact caused by the truncation of the Hilbert space. We know this because it gives a positive second derivative of the mass exponent, while this derivative must always be negative~\cite{Janssen1994}. In fact, the point where the curvature of the $\tau_{q}$ becomes positive could be used as an alternative tool to determine the convergence of the wave function. Our current convergence criterion for the wave functions is established for $q=1$ to ensure that most of the state is contained within the energy domain of interest~\cite{Bastarrachea2014PSa,Bastarrachea2014PSb}. It is impossible to guarantee the convergence of the whole wave function unless we set the truncation to infinity. This problem becomes more important for highly extended states. For low energy regimes, such as $\epsilon_{0}=-1.8$ and $-1.5$, the small phase space available is another problem that one needs to keep in mind. In this case, a finite-size effect arises, turning the curvature of $\tau_{q}$ from negative to positive. This effect is a result of the artificial truncation of low-energy states due to the Hilbert space being bounded from below, as can be seen in the PDoS$_{q}$ in Figs.~\ref{fig:1}~(a2)-\ref{fig:1}~(a4). This finite-size effect can be removed by increasing $j$, but this requires even larger values of $n_{\text{max}}$ to ensure overall convergence. We must discard any result that gives a positive curvature of $\tau_{q}$ when fitting the curves to get $D_q$.

\subsubsection{Magenta circle in Fig.~\ref{fig:1}~(a1)}

We now focus on the coherent state centered at the magenta point in the phase space of Fig.~\ref{fig:1}~(a1). The state is inside a regular region but surrounded by regular orbits resulting from a nonlinear resonance that appears in the Dicke Hamiltonian between the two normal modes at low energies~\cite{Reichl1987,Relano2016,Bastarrachea2017}. Like the case of the green circle, for $q=1$ and $q=2$, the PDoS$_{q}$ shows the dominance of a single Gaussian, while for $q=1/2$, more Gaussians appear. For this state, we see that $\tau_q$ is nonlinear not only for $q<1$, as before, but also for $q>1$. The slope of the curve for $\tau_{q}$ is close to $D_{q}^{(c)}=1/3$, but deviates from it in both limits of $q$. 

We compare the curve for $\tau_q$ versus $q$ with different fittings in Fig.~\ref{fig:4}~(c) for $q<1$ and in Fig.~\ref{fig:4}~(d) for $q>1$. Unlike the green-circle coherent state in Fig.~\ref{fig:4}~(a), for the magenta-circle coherent state, the quadratic fit for $q<1$ [Fig.~\ref{fig:4}~(c)] is better than the one that goes as $q^{1/2}$. As we turn to $q>1$, we see that for $q \gg 1$, a linear fitting seems to be the best one, although for $q \gtrsim 1$, the quadratic fitting describes well the nonlinearity. While both coherent states marked by the green and magenta circles are in a regular region, the magenta one is surrounded by orbits associated with a nonlinear resonance. This may impact the structure of the coherent states and be associated with the multifractal behavior for $q>1$.

\subsection{Mixture of chaos and regularity }

By increasing the energy, we introduce a mixture of chaos and regularity, as shown in Fig.~\ref{fig:1}~(b1) for $\epsilon_{0}=-1.5$. For our analysis, we choose three coherent states with this energy and centered at the points marked with diamonds in Fig.~\ref{fig:1}~(b1). The point indicated by a green diamond is away from the three major regular islands and inside an emergent sea of chaos. The coherent state associated with this point shows a behavior similar to that of the yellow circle with $\epsilon_{0}=-1.8$ in Fig.~\ref{fig:1}~(a3). The PDoS$_{q}$ exhibits clustered Gaussians, and $\tau_{q}$ is almost linear with a slope very close to $D_{q}^{(r)}=1$, which indicates that this coherent state is ergodic.

The magenta diamond in Fig.~\ref{fig:1}~(b1) is located in a mixed region. Compared with the green-diamond coherent state, the components of the PDoS$_{q}$ for the magenta-diamond state are less concentrated around the energy of the surface. Even though $\tau_q$ presents a linear behavior [see Fig.~\ref{fig:1}~(b4)], indicating that the state is monofractal, the extracted value of $D_{q}$ is between those for a random and a regular state. 

The coherent state indicated with the yellow diamond in Fig.~\ref{fig:1}~(b1) lies in a similar position to the magenta circle in Fig.~\ref{fig:1}~(a1), i.e., at the center of the regular island of the nonlinear resonance orbits. While it exhibits nonlinear behavior for $q<1$ [see Fig.~\ref{fig:1}~(b3)], it shows an almost perfect linear behavior for $q>1$ with $D_{q}^{(c)}=1/3$. Finding linear behaviors like this allows us to use $D_{q}$ to identify regular regions in phase space. 

\subsection{Predominance of chaos}

We now explore a higher energy ($\epsilon_{0}=-1.1$) in Fig.~\ref{fig:2}, where chaos is dominant, but some regions of regularity still exist. The first coherent state that we select is amidst the chaotic sea and is marked with a green square in Fig.~\ref{fig:2}~(a1). The components of the PDoS$_{q}$ next to Fig.~\ref{fig:2}~(a2) are randomly distributed, but some regular structures can still be identified. Hence, we obtain a slope for the $\tau_q$ curve that is between $D_{q}^{(c)}$ and $D_{q}^{(r)}$, similarly to the coherent state marked with the magenta diamond in Fig.~\ref{fig:1}~(b4). 

In contrast to the above, the coherent state marked with the yellow square in Fig.~\ref{fig:2}~(a3) is fully extended. This is not obvious from the PDoS$_{q}$, as it seems to be shaped by two Gaussians for $q=1$. Instead, when we go to $q<1$, all the Gaussians seem to have comparable weight. The ergodic nature of the state gets determined from the slope of the $\tau_{q}$, which is constant and equal to $D_{q}^{(r)}=1$. 

Moving to larger values of $j_{z}$ in the Poincar\'e section of Fig.~\ref{fig:2}~(a1), we reach the magenta square, which is at a regular island. The slope of the curve for $\tau_q$ in Fig.~\ref{fig:2}~(a4) is close to $D_{q}^{(c)}=1/3$, confirming that the state is regular. This means that by studying the behavior of the mass exponents, we can locate regular regions even in very chaotic domains.

\begin{figure}
\centering
\begin{tabular}{c}
\includegraphics[width=0.5\textwidth]{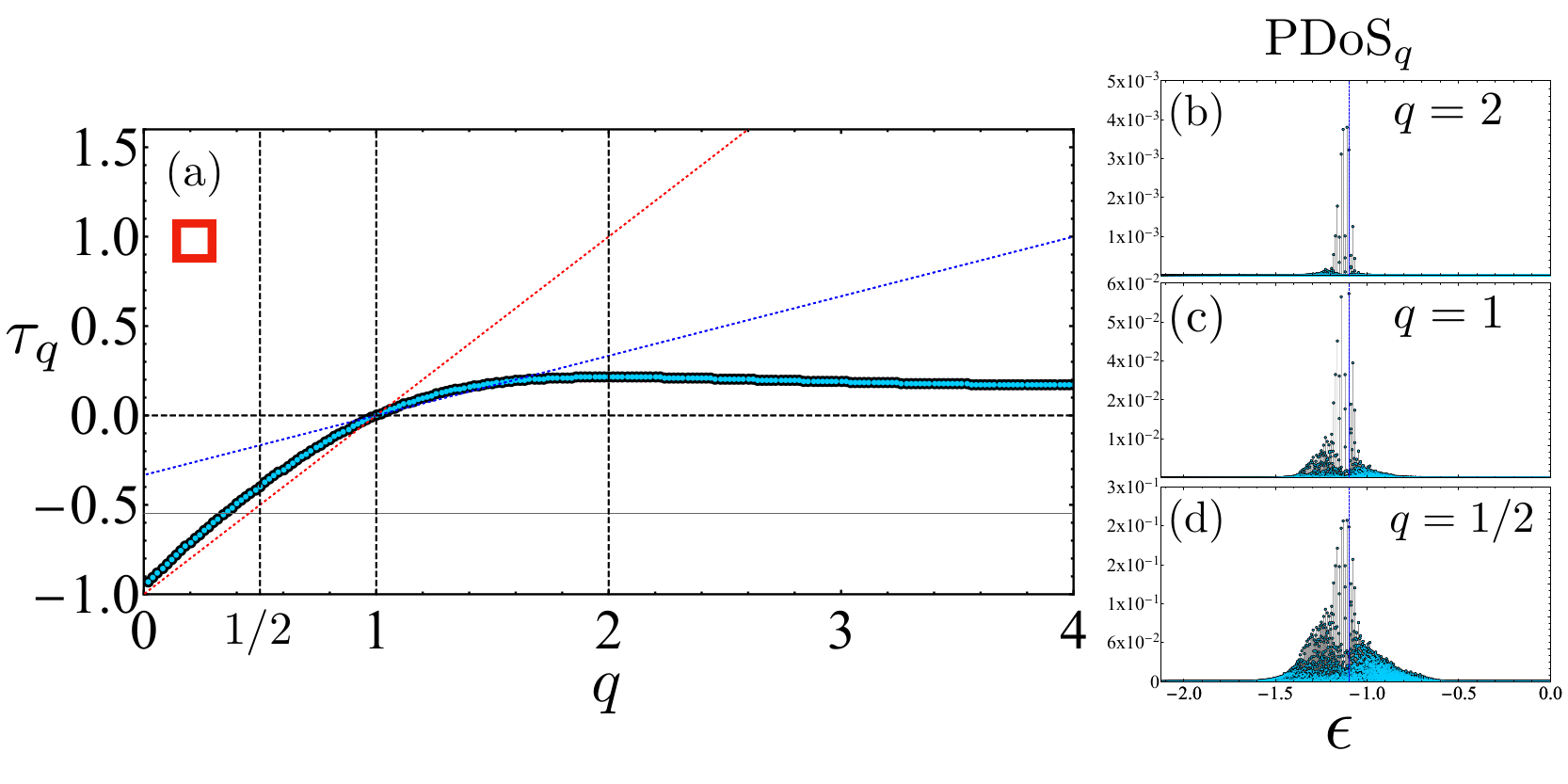}
\end{tabular} 
\caption{(a) Mass exponents $\tau_q$ as a function of $q$ for the coherent state marked by the white square encircled in red in Fig.~\ref{fig:2}~(a1) at $\epsilon=-1.1$, and PDoS$_{q}$ for (b) $q=2$, (c) $q=1$, and (d) $q=1/2$.}
\label{fig:5}
\end{figure}

In Fig.~\ref{fig:2}~(a1), we identify another regular coherent state with special behavior. It is located in the other stability island in Fig.~\ref{fig:2}~(a1) and is marked with a white square encircled in red. The behavior of $\tau_{q}$ for this coherent state is shown in Fig.~\ref{fig:5}~(a). It presents a highly nonlinear behavior for $q<2$, while $\tau_{q}$ becomes nearly flat for $q>2$. This indicates that this state has an effective dimension of $D_{\infty}\simeq 0$. This is confirmed by looking at the PDoS$_{q}$ in Fig.~\ref{fig:5}~(b). We see in Figs.~\ref{fig:5}~(b)-\ref{fig:5}~(d) that as $q$ increases from $q=1/2$ to $q=2$, the number of components relevant to describe the state decreases quickly, in agreement with the idea that when $D_{q}=0$, we have a state with zero measure in the Hilbert space. However, for $q<1$, we see in Fig.~\ref{fig:5}~(a) that $D_{q}$ follows the behavior of an ergodic state with $D_{q}^{(r)}$, which is also suggested by the PDoS$_{q}$ in Fig.~\ref{fig:5}~(d). This state is of particular interest because it exhibits a highly multifractal behavior. For $q<1$, it behaves as a Gaussian state, while for $q>1$, $D_{\infty}$ tends to 0. Although the phase space might look qualitatively similar in the Poincaré section, the multifractal method unveils quantitative differences between this coherent state and that  marked by the magenta square at the same energy but in another regular zone [see Fig.~\ref{fig:2}~(a4)]. Because of the resolution limits imposed by the numerical diagonalization, a supplementary approach as a dynamical tool, such as the survival probability~\cite{Lerma2018} or out-of-time ordered correlators~\cite{Hashimoto2017}, could help to understand the details of the participating structures and unveil the reasons behind the observed multifractal behavior.

In Fig.~\ref{fig:2}~(b1), we consider an even higher energy ($\epsilon_{0}=-0.5$), where one expects ergodic behavior~\cite{Cejnar2021,Chavez2016}. The three selected coherent states are centered at the points labeled by stars. As shown in Figs.~\ref{fig:2}~(b2)-\ref{fig:2}~(b4), they are not multifractal but just single fractals with an approximated dimension close to $D_{q}^{(r)}=1$, indicating ergodicity. The different shape of the phase space happens because the two wells at low energies in the energy surface of the Dicke model merge together~\cite{Bastarrachea2014a}.

\section{Fractal dimension over energy surfaces}
\label{sec:5}  

Now that the detailed study of representative coherent states is complete, we proceed to build a general characterization of the energy surfaces through the information stored in the mass exponents. To this end, we fit the curve for $\tau_q$ versus $q$ with the parabolic approximation $\tau_{q}=\mathcal{D}^{(0)}+\mathcal{D}^{(1)}(q-1)+\mathcal{D}^{(2)}q^{2}$. This way, we can extract the linear part of the fit, $\mathcal{D}^{(1)}$, and use it as a probe to determine whether the state is ergodic ($\mathcal{D}^{(1)}=D_{q}^{(r)}=1$), regular ($\mathcal{D}^{(1)}=D_{q}^{(c)}=1/3$), localized  ($\mathcal{D}^{(1)}=D_{q}=0$), or anything in between. We can also obtain the quadratic part, $\mathcal{D}^{(2)}$, and use it as a first-order measure of the nonlinear behavior of the state, hence the presence of multifractality. For our analysis, we study the sample of points over the Poincar\'e surface of sections with $\phi_{0}=0$ marked in Figs.~\ref{fig:1} and~\ref{fig:2}, panels~(a1) and (b1). 

The linear part of the fit $\mathcal{D}^{(1)}$ is used to approximate the linear part of the generalized dimension $D$ fitted over the $q>1$ domain. We investigate $\mathcal{D}^{(2)}q^{2}$ for both $q<1$ and $q>1$ to estimate the departure from the linear behavior quantified by $\Delta$.

\begin{figure}
\centering
\begin{tabular}{c}
\includegraphics[width=0.5\textwidth]{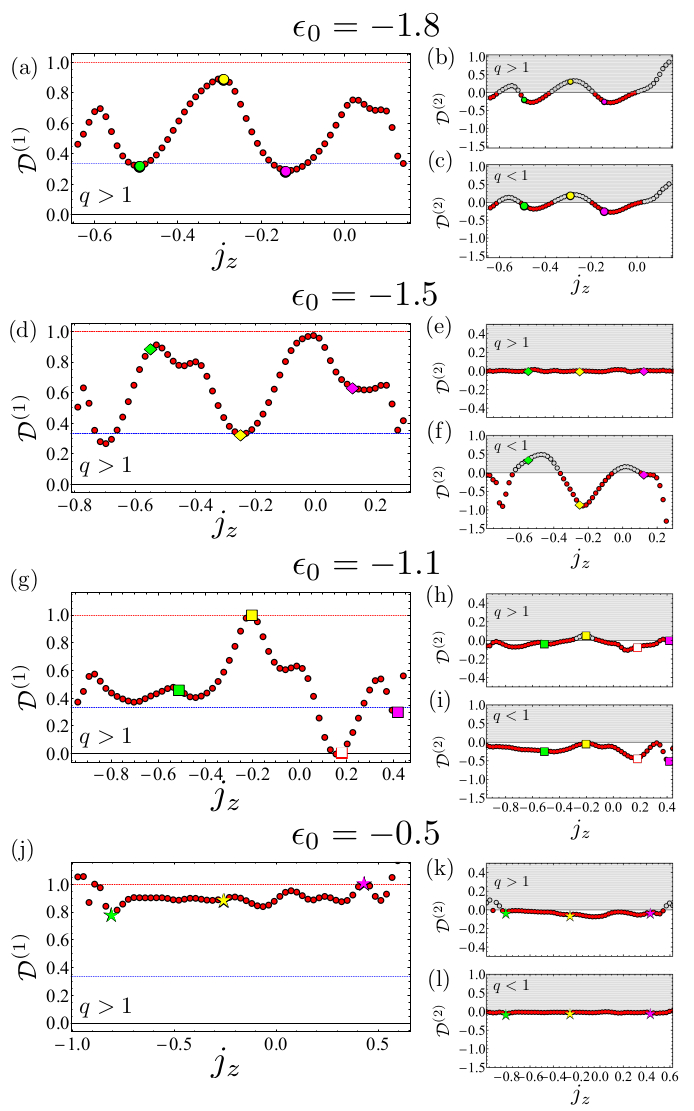}
\end{tabular} 
\caption{Linear approximation to the generalized dimension $\mathcal{D}^{(1)}$ as a function of $j_{z}$ with $\phi=0$ over the energy surface for (a) $\epsilon_{0}=-1.8$, (d) $\epsilon_{0}=-1.5$, (g) $\epsilon_{0}=-1.1$, and (j) $\epsilon_{0}=-0.5$. Parabolic approximation to the anomalous fractal dimensions $\mathcal{D}^{(2)}$ over the same energy surfaces for $q>1$ [(b, e, h, k)] and $q<1$ [(c, f, i, l)]. The region in gray indicates the points that should not be considered, as $\mathcal{D}^{(2)}$ becomes positive.}
\label{fig:6}
\end{figure}

\subsection{Ergodic versus multifractal coherent states}

We first use the parabolic approximation to fit the mass exponents for $q>1$, in the region $q\in[1,2]$ and focus on the analysis of the linear approximation characterized by $\mathcal{D}^{(1)}$. Under the weak multifractality approximation, as discussed in the context of Eq.~\eqref{Eq:Deltaq}, the coefficient to the linear part should formally be $\mathcal{D}^{(1)}=D+\Delta$. However, as the parabolic approximation does not hold systematically over the whole phase space, we work under the approximation that $\mathcal{D}^{(1)}\simeq D$ for $q>1$. 

Our results for $\mathcal{D}^{(1)}$ as a function of $j_z$ are shown in  Figs.~\ref{fig:6}~(a), \ref{fig:6}~(d), \ref{fig:6}~(g), and \ref{fig:6}~(j) for the energy surfaces studied in Figs.~\ref{fig:1} and~\ref{fig:2}. For the surface with $\epsilon_{0}=-1.8$ [Fig.~\ref{fig:6}~(a)], where regularity dominates, $\mathcal{D}^{(1)}$ locates the regular regions around $j_{z}\sim -0.5$, $j_{z}\sim - 0.2$, and $j_{z}\sim 0.1$, where $\mathcal{D}^{(1)}\sim 1/3$. Notice that the lowest value of $\mathcal{D}^{(1)}$ for this energy surface with $\epsilon_{0}=-1.8$ is $D_{q}^{(c)}$, so there are no states with features of localization, that is, $\mathcal{D}^{(1)}$ is never close to zero. The value of $\mathcal{D}^{(1)}$ in Fig.~\ref{fig:6}~(a) also detects the emergence of chaos. Even though it does not reach $D_{q}^{(r)}$, it does capture significantly extended states around $j_z \sim -0.3$, where the stochastic layer is developing. 

In the mixed region at $\epsilon_{0}=-1.5$ [Fig.~\ref{fig:6}~(d)],  $\mathcal{D}^{(1)}$ again detects regular and chaotic regions. For $j_{z}\sim 0.0$, $\mathcal{D}^{(1)}$ now reaches $D_{q}^{(r)}$, indicating the presence of ergodic states and thus of fully chaotic regions. 

The most interesting results appear at the energy surface $\epsilon_{0}=-1.1$ [Fig.~\ref{fig:6}~(g)], where chaos has developed, and only a few regular domains survive in phase space. This figure makes it evident that the linear approximation to the fractal dimension $\mathcal{D}^{(1)}$ provides more detail about the phase space than the Poincar\'e section.  While the latter suggests that most of the phase space is fully chaotic, $\mathcal{D}^{(1)}$ shows that only the states around $j_z\sim-0.2$ are fully extended, having $\mathcal{D}^{(1)}\sim D_{q}^{(r)}=1$. As we move away from this point, towards larger values of $j_{z}$, there is an abrupt decrease in the values $\mathcal{D}^{(1)}$ until it reaches $\mathcal{D}^{(1)}=0$ around $j_{z}\simeq 0.1$, indicating that we have found a state with localized features for $q>1$. If instead, we move away from $j_z\sim-0.2$, towards smaller values of $j_{z}$, then we see that for the coherent states with $j_z<-0.4$, $\mathcal{D}^{(1)}$ has values similar to those seen for surfaces with low energies, getting close to $D_{q}^{(c)}=1/3$. This implies that energy surfaces with qualitatively different phase space structures may actually have similar values of the approximated fractal dimension.

For the energy surface at high energy, $\epsilon_{0}=-0.5$ in Fig.~\ref{fig:6}~(j), we expect ergodic behavior. Indeed, over the whole energy surface, $\mathcal{D}^{(1)}\sim D_{q}^{(r)}=1$. This analysis demonstrates that $D=\mathcal{D}^{(1)}$, despite belonging to the quantum world, is a good quantity to probe chaos and regularity in phase space, and to uncover its structures.

\subsection{Multifractal versus fractal coherent states}

The approximated linear fractal dimension, $D\simeq\mathcal{D}^{(1)}$, allows for the quantitative classification of the coherent states into ergodic, regular, or an intermediate fractal case. However, it only reveals single fractal behavior. Extended states that are neither ergodic ($D_{q}=1$) nor regular ($D_{q}=1/3$) can be mono- or multifractal. To quantify multifractality and distinguish between mono- and multifractal states, we need to study the deviations from linear behavior. For this, we look at the quadratic order of the fitting curve for the mass exponents,  $\tau_{q}=\mathcal{D}^{(0)}+\mathcal{D}^{(1)}(q-1)+\mathcal{D}^{(2)}q^{2}$, for both $1\leq q\leq 2$ and $0.3 \leq q \leq 1$. If the value of the quadratic term in the fractal dimension $\mathcal{D}^{(2)}$ is different from zero and negative, then the mass exponent has a nonlinear behavior of the order of weak multifractality. It is worth emphasizing that the quadratic fit is not general, as the anomalous fractal dimension $\Delta_{q}$ could be an arbitrary function of $q$ (under the curvature constraints for $\tau_{q}$). Yet, we choose the quadratic fitting as a general way to detect nonlinearity, so when the behavior is parabolic $\mathcal{D}^{(2)}\simeq\Delta$.  

At the energy surfaces $\epsilon_{0}=-1.8$ [Figs.~\ref{fig:6}~(b) and \ref{fig:6}~(c)] and $\epsilon_{0}=-1.1$ [Figs.~\ref{fig:6}~(h) and \ref{fig:6}~(i)], we observe that for $q>1$, most states have $\mathcal{D}^{(2)}\sim 0$, except those around the regular islands. In this case, there is a slight decrease in the value of $\mathcal{D}^{(2)}$ indicating the multifractal nature of those states. The same happens for $q<1$, where $\mathcal{D}^{(2)}$ signals nonlinear behavior and weak multifractality around the regular islands. Notice that $\mathcal{D}^{(2)}$ becomes positive for some states, typically the most extended ones. This is an artifact caused by the combination of the already mentioned finite-size effects at low energies and the limitations of the wave-function convergence. Since $\Delta$ must always be negative, any value of $\mathcal{D}^{(2)}$ different from zero but positive must be discarded. The region where the points need to be discarded is indicated in gray in Figs.~\ref{fig:6}~(b), \ref{fig:6}~(c), \ref{fig:6}~(e), \ref{fig:6}~(f), \ref{fig:6}~(h), \ref{fig:6}~(i), \ref{fig:6}~(k), and \ref{fig:6}~(l). 

Nonlinearity is more visible for $q<1$ when $\epsilon_{0}=-1.5$ [Fig.~\ref{fig:6}~(f)]. One sees large oscillations in the values of $\mathcal{D}^{(2)}$. In contrast, for $\epsilon_0=-0.5$ in Figs.~\ref{fig:6}~(k) and \ref{fig:6}~(l), we confirm that the system mainly comprises ergodic states. 

In general, we conjecture that structural changes in phase space from an extended region to a very localized region, as it occurs for the $D^{(1)}$ at $\epsilon_{0}=-1.5$ [Fig.~\ref{fig:6}~(g)], produce significant multifractal behavior, which is identifiable for $\mathcal{D}^{(2)}$ and $q>1$ [Fig.~\ref{fig:6}~(i)].
From the three energy cases in Fig.~\ref{fig:6}, $\epsilon_{0}=-1.8,-1.5,-1.1$, it seems that the coherent states have to gain a multifractal character across the phase space as they transit from ergodicity to regularity or $D=0$.

A systematic study of nonlinearities as one moves through phase space is complicated. For example, we showed in Fig.~\ref{fig:4}~(a)  that the coherent state marked by the green circle in Fig.~\ref{fig:1}~(a1) at $\epsilon_{0}=-1.8$ is better fitted by a curve of the form $\tau_{q}=\mathcal{D}^{(0)}+\mathcal{D}^{(1)}(q-1)+\mathcal{D}^{(1/2)}q^{1/2}$ for $q<1$, so a detailed analysis to determine the true multifractal nature of a coherent state must be done individually. Nevertheless, our analysis in this section has demonstrated that the approximation $\mathcal{D}^{(2)}\simeq\Delta$ is enough for a coarse-grained distinction between monofractal and multifractal states.

\section{Conclusions and perspectives}
\label{sec:6}  

We have analyzed the fractal properties of coherent states projected in the energy eigenbasis of the Dicke Hamiltonian. The motivation to explore multifractality in this model comes from its experimental accessibility, its well-defined classical limit, and the fact that it represents many-body systems with collective interactions and only two degrees of freedom, which simplifies the description. However, the model has an unbounded Hilbert space, which makes the study of multifractality based on scaling analysis challenging. To obtain the spectrum numerically, one must choose a cut-off of the bosonic space large enough to ensure convergence of the high-energy eigenstates. We used cut-off values that were large enough to ensure the normalization of the wave functions and the accurate computation of the generalized inverse participation ratio with $q>1$, but nonconverged components still impact the multifractal analysis when one goes to small values of $q$. We have circumvented this problem by restricting our study to certain ranges of $q$ values.

Although the multifractal analysis has been used before as a probe of ergodicity, here our aim was to employ it as a coarse-grained quantitative tool to capture changes in the structures in a mixed phase space. The multifractal analysis of coherent states reveals details of the rich phase space of the Dicke model and provides a quantitative picture of its structure. We have obtained two main results by studying the mass exponents $\tau_q$. 

First, by fitting the curves of $\tau_q$ versus $q$, we verified that the approximation to the linear generalized dimension, $\mathcal{D}^{(1)}\simeq D$, is a valuable tool to distinguish regular from chaotic regions, similar to what was done for the kicked top model in Ref.~\cite{Wang2021} using $D_1$, $D_2$ and $D_\infty$, and for the Dicke model using $D_2$~\cite{Bastarrachea2016,Bastarrachea2017}. The value of $D$ operates as a sensitive probe of the phase space that allows for the identification of chaotic states in regular regions and regular states associated with islands of stability in the chaotic regime. 

Second, we showed that the parabolic correction, $\mathcal{D}^{(2)}\simeq \Delta$, to the linear fit of the mass exponents works as a probe of nonlinearity and thus reveals the presence of multifractal coherent states. We emphasize that states whose linear fractal dimension $D$ is constant and between $D_{q}^{(r)}$ (ergodic) and $D_{q}=0$ (fully localized) are nonergodic extended states that are fractal, but not necessarily multifractal. For multifractality to hold, $D_q$ needs to show a nonlinear behavior with $q$, which can be detected with the analysis of $\mathcal{D}^{(2)}$. 

The multifractal analysis unveils the presence of nonlinearities in the quantum states. From the quantum-classical correspondence, we know that the coherent state samples the vicinity of the phase space around its center, so the presence of multifractality must be a reflection of neighboring hidden structures in the phase space, hinted by the examination of the PDoS$_{q}$. Multifractality is thus related to dramatic changes in phase space and to the simultaneous participation of different phase space structures.  
 
Moreover, multifractality in the sense of nonergodic extended states was recently studied in the Tavis-Cummings model, an integrable limit of the Dicke model~\cite{Mattiotti2023}. Alongside our results in the regular region of the Dicke model and those in Ref.~\cite{Atas2012}, it is clear that regular systems can also exhibit multifractality. We leave for a future work, the analysis of how the multifractality of coherent states over the energy eigenbasis gets manifested at the level of classical dynamics. Likewise, recently, there has been an active interest in developing dynamic quantifiers of chaos, including the survival probability for a nonstationary state after a quantum quench~\cite{Lerma2018,Lerma2019,Villasenor2020,Zarate-Herrada2023}, or the different out-of-time-ordered correlators (OTOCs)~\cite{Chavez2019,Tiwari2023}.
Most of these indicators require information on the eigenfunctions of the Hamiltonian to calculate the dynamics. We deem the multifractal analysis as a complementary approach to these dynamical indicators, and future work will be devoted to understanding more about the phase space structures participating in each coherent state. We hope that the results of our work will motivate the study of multifractality in other systems with bounded and unbounded Hilbert spaces.

Measuring experimental signatures of multifractality in quantum systems has remained elusive to date. Some progress has been made with cold atoms~\cite{Lemarie2010,Sagi2012,Lopez2013}, disordered conductors~\cite{Richardella2010}, and open three-dimensional elastic networks~\cite{Faez2009}. Three conditions are necessary to detect quantum multifractal features: the ability to measure the wave function in the chosen basis, scalability, and robustness against perturbations~\cite{Dubertrand2014,Dubertrand2015}. For our scheme of coherent states in the Dicke model, we believe that ion trap platforms may offer a good route to explore multifractal signatures~\cite{Safavi2018,Cohn2018,Lewis-Swan2019}, because of its scalability~\cite{Bohnet2016} and the possibility of exploring a wide range of Hamiltonian parameters~\cite{Lewis-Swan2021}. In these experiments, the spin degree of freedom is encoded in two internal states of the ions, and the boson is realized through the collective center-of-mass motional mode of the ion crystal. While it is currently possible to prepare the system in low-lying states, the challenge would be to prepare it in a high-lying energy coherent state. Another possible approach to extract multifractal information via quantum simulation is to employ the quantum wavelet transform, given that the Dicke model and the coherent states can be efficiently simulated on a quantum computer~\cite{Garcia-Mata2009}.

\begin{acknowledgments} 
We acknowledge insightful comments by S. Pilatowsky-Cameo, T. Lezama Mergold-Love, and M. de J. G\'onzalez Mart\'inez. We acknowledge also the support of the Computation Center - ICN, in particular to Enrique Palacios, Luciano D\'iaz, and Eduardo Murrieta. This research was partially funded by the PRODEP Project No. 12313674, the DGAPA-UNAM Project No. IN109523. D.V. acknowledges financial support from the postdoctoral fellowship program DGAPA-UNAM. J.C.C. was funded by the United States NSF CCI Grant (Award No. 2124511). L.F.S. was supported by the United States NSF, Grant No. DMR-1936006.
\end{acknowledgments}

\appendix

\section{Scaling of typical states}
\label{app:1}

In this Appendix we calculate the two bounds for the scaling of the  $\text{IPR}_{q}$ discussed in the main text. 

\subsection{Scaling of random state with Gaussian profile}

We consider a state that has a Gaussian distribution over the eigenbasis with random coefficients
\begin{gather}
|c_{k}^{(r)}|^{2}=\frac{r_{k}}{\sum_{k}\exp\left(-\frac{(E_{k}-\bar{E})}{2\sigma_{r}^{2}}\right)}\,\exp\left(-\frac{(E_{k}-\bar{E})}{2\sigma_{r}^{2}}\right),
\end{gather}
where $r_{k}$ is a random number, $\bar{E}$ is the average energy of the state, and the variance $\sigma_{r}$ is chosen arbitrarily. When the dimension of the system grows, i.e., in the limit $j>>1$, we approximate the sum by an integral
\begin{gather}
\sum_{k}\exp\left(-\frac{(E_{k}-\bar{E})^2}{2\sigma_{r}^{2}}\right)=\frac{1}{\Delta E}\sum_{k},\exp\left(-\frac{(E_{k}-\bar{E})^2}{2\sigma_{r}^{2}}\right)\Delta E\\ \nonumber \simeq\bar{\nu}\int\exp\left(-\frac{(E-\bar{E})^2}{2\sigma_{r}^{2}}\right)\, dE=\bar{\nu}\sqrt{2\pi}\sigma_{r},
\end{gather}
where $\bar{\nu}$ is the average density of states around the energy window. Therefore, 
\begin{equation}
|c_{k}^{(r)}|^{2}=\frac{r_{k}}{\sqrt{2\pi}\sigma_{r}\bar{\nu}}\,\exp\left(-\frac{(E_{k}-\bar{E})^2}{2\sigma_{r}^{2}}\right).
\end{equation}
Now, we calculate the $\text{IPR}_{q}$, 
\begin{gather}
\nonumber
\text{IPR}_{q}^{(r)}=\sum_{k}|c_{k}^{(r)}|^{2q}=\sum_{k}\left(\frac{r_{k}}{\sqrt{2\pi}\sigma_{r}\bar{\nu}}\,\exp\left(-\frac{(E_{k}-\bar{E})^2}{2\sigma_{r}^{2}}\right)\right)^{q} \\ \nonumber
= \frac{1}{(\sqrt{2\pi}\sigma_{r}\bar{\nu})^{q}}\sum_{k}r_{k}^{q}\,\exp\left(-\frac{(E_{k}-\bar{E})^2}{2\sigma_{r}^{2}}\,q\right) \\ \nonumber
= \frac{\bar{\nu}}{(\sqrt{2\pi}\sigma_{r}\bar{\nu})^{q}}\int r^{q}(E)\,\exp\left(-\frac{(E-\bar{E})^2}{2\sigma_{r}^{2}}q\right)\,dE \\ \nonumber
= \frac{1}{(\sqrt{2\pi}\sigma_{r}\bar{\nu})^{q}}\bar{\nu}\bar{r^{q}}\sqrt{\frac{2\pi}{q}}\sigma_{r} \nonumber\\
= \left[\frac{(2\pi)^{(1-q)}}{q}\right]^{1/2} \bar{r^{q}}(\sigma_{r}\bar{\nu})^{1-q}.
\label{eq:IPRrand}
\end{gather}
The random state samples all the energies present in the energy window, and $\bar{\nu}\sim\nu_{0} j$~\cite{Bastarrachea2014a}. Additionally, we chose $\sigma_{r}$ corresponding to a coherent states, which scales as $\sigma_{r}\sim j^{1/2}$~\cite{Schliemann2015}. Then, the scaling of the generalized inverse participation ratios, in this case, is
\begin{equation}
\text{IPR}_{q}^{(r)}\propto j^{(3/2)(1-q)}.
\end{equation}
Then, for a random state with a Gaussian profile, we have a similar result as that of an ergodic (extended) case, where $D_q=1$ for all values of $q$.

\subsection{Scaling of a regular state}

Let us now consider the limiting case when a coherent state is centered in a point in phase space just in the center of an island of regularity. As shown in Ref.~\cite{Lerma2018}, in this case the coherent state spanned over the eigenbasis has only main contributions from a subset of energy levels, which form a sequence of nearly equally spaced energies, $\Delta E_k^{(\text{seq})}=E_{k+1}^{(\text{seq})}-E_{k}^{(\text{seq})}\xrightarrow{j\rightarrow\infty} \omega_{\text{cl}}$, where $\omega_{\text{cl}}$ is a finite value, corresponding to the classical frequency of the classical mode activated in the center of the stability island. The squared magnitude of the sequence coefficients of this state are also described by a Gaussian profile, 
\begin{equation}
|c_{k}^{(\text{seq})}|^{2}=\frac{1}{\sum_{k}\exp\left(-\frac{E_{k}^{(\text{seq})}-\bar{E}}{2\sigma_{c}^{2}}\right)}\,\exp\left(-\frac{E_{k}^{(\text{seq})}-\bar{E}}{2\sigma_{c}^{2}}\right),
\end{equation}
where $\sigma_{c}$ is the variance of the Gaussian profile over the single contributing eigenenergy sequence. By following a similar procedure to that yielding  Eq.~\eqref{eq:IPRrand}, we obtain now
\begin{equation}
\text{IPR}_{q}^{(c)}=\sqrt{\frac{(2\pi)^{(1-q)}}{q}}(\sigma_{c}\bar{\nu}_{c})^{1-q}.
\end{equation}
However, unlike the previous case,  $\bar{\nu}_c$ is not the total density of states, but the density of states of the participating sequence of eigenstates, which is given by $\bar{\nu}_c=1/\Delta E_k^{(\text{seq})}$ and tends to a finite value for large $j$, 
$$\bar{\nu}_c\xrightarrow{j\rightarrow\infty}\frac{1}{ \omega_{\text{cl}}}.$$
However, as before, $\sigma_{c}\sim j^{1/2}$ because it comes from a coherent state~\cite{Schliemann2015,Lerma2018}. In this way,  the scaling of the $\text{IPR}_{q}^{(c)}$ is, 
\begin{equation}
\text{IPR}_{q}^{(c)}\sim j^{(1/2)(1-q)}=\left[j^{(3/2)}\right]^{1/3(1-q)}.
\end{equation}
Therefore, for this ideal case where the components out of the main sequence are strictly zero, we get $D_q^{(c)}=1/3$ for the generalized dimension. However, for the coherent states in the center of the island of stability considered in the main text, the components out of the main sequences are much smaller, but not zero. As a consequence, in the limit of $q\rightarrow 0$ the generalized dimension should tend to the maximum value in the space, i.e., $D_{q\rightarrow 0}=1$. This explains why some coherent states around stable points have curvature for $q\ll 1$. Also, this result shows that a Gaussian state possesses an intrinsic multifractal effect as the $D_{q}$ must go from $1/3$ to $1$ as $q$ decreases. This is not a finite-size effect.

\section{Increasing eigenstate convergence}
\label{app:3}

\begin{figure}[h!]
\centering
\begin{tabular}{c}
\includegraphics[width=0.4\textwidth]{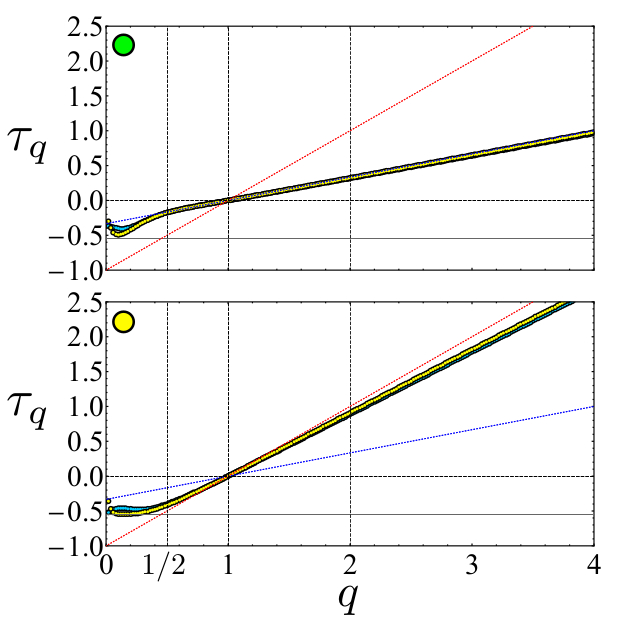}
\end{tabular} 
\caption{Comparison between the mass exponents $\tau_q$ for the respective coherent states in Figs.~\ref{fig:1}~(a2) and \ref{fig:1}~(a3) at $\epsilon_{0}=-1.8$. The data was adjusted for $j=5$ to $100$ and two values of the cut-off: $n_{\text{max}}=250$ (yellow dots) and $n_{\text{max}}=300$ (blue dots).}
\label{fig:7}
\end{figure}

In Fig.~\ref{fig:7}, we show the results that we obtain by decreasing the cut-off over the bosonic Hilbert space of the Dicke model from the value used in the main text $n_{\text{max}}=300$ (blue dots) to $n_{\text{max}}=250$ (yellow dots). We employ the same representative points in phase space that are used in Figs.~\ref{fig:1}~(a2) and \ref{fig:1}~(a3) at the energy surface to $\epsilon_{0}=-1.8$. We observe that the results are robust and only slightly deviate for large values of $q$. The main difference appears for $q<1$. For $n_{\text{max}}=300$, the change in the slope for $q<1$ signaling the limits on the convergence of the wave function occurs for smaller values of $q$ than for $n_{\text{max}}=250$, although this is hardly noticeable in the figure. The distinction between the two curves becomes visible only for very small values of $q$.

\bibliography{MF_Notes}

\end{document}